**O. V. KOMPANIIETS**[1], Junior Researcher
https://orcid.org/0000-0002-8184-6520
E-mail: kompaniets@mao.kiev.ua
**I. B. VAVILOVA**[1], Head of Department, Dr. Sci. Hab. in Phys. Math., Prof.
https://orcid.org/0000-0002-5343-1408
**O. M. KUKHAR**[1], Junior Researcher, Post-graduate Student
https://orcid.org/0009-0000-7042-9214
**D. V. DOBRYCHEVA**[1], Senior Researcher, Cand. Sci. in Phys. Math.
https://orcid.org/0000-0001-5557-3453
**P. N. FEDOROV**[2], Head of Laboratory, Dr. Sci. Hab. in Phys. Math.
https://orcid.org/0000-0001-6745-3731
**A. M. DMYTRENKO**[1], Senior Researcher, PhD
https://orcid.org/0000-0001-9380-6641
**V. P. KHRAMTSOV**[1], Software Developer
https://orcid.org/0000-0003-1744-7071
**O. SERGIJENKO**[1], Senior Researcher, Cand. Sci. in Phys. Math.
https://orcid.org/0000-0002-9212-7118
**A. A. VASYLENKO**[1], Senior Researcher, Cand. Sci. in Phys. Math.
https://orcid.org/0000-0001-7142-6576

[1] Main Astronomical Observatory of the National Academy of Sciences of Ukraine
27 Akademik Zabolotnyi Str., Kyiv, 03143 Ukraine
[2] Institute of Astronomy of V. N. Karazin Kharkiv National University
4 Svoboda Sq., Kharkiv, 61022 Ukraine

# MILKY WAY GALAXY-ANALOGS AND ISOLATED GALAXIES WITH BARS: ENVIRONMENTAL DENSITY IN THE LOCAL VOLUME

*The environmental density of galaxies within the cosmic web provides insight into their 3D locations in filaments, voids, groups, and clusters of the large-scale structure of the Universe. This parameter reflects the distribution of baryonic matter and the influence of dark matter halos on galaxy evolution. Understanding environmental density is crucial for identifying the external physical processes — such as feedback from supernovae and active galactic nuclei, tidal interactions, ram-pressure stripping, and large-scale matter flows — that shape galaxies beyond their internal properties. In this article, we focus on the isolated galaxies with bars as the ensemble of galaxies for which the Milky Way galaxy-analogs belong. To obtain local environmental density parameters and verify the isolation criterion ($|\Delta v| \le 500$ km/s), we developed a Python-based pipeline operating in two redshift regimes: low ($z_0 < 0.02$) and high ($z_0 \ge 0.02$). Local densities Σ were estimated using both k-nearest neighbor and Voronoi tessellation methods and classified as void ($\Sigma < 0.05$), filament ($0.05 \le \Sigma < 0.5$), group ($0.5 \le \Sigma < 2.0$), and cluster ($\Sigma \ge 2.0$).*

*Our sample of 311 isolated barred galaxies from the 2MIG catalog, supplemented by Milky Way-analog systems ($z < 0.07$), covers the northern sky. We find 157 galaxies in voids, 84 in filaments, 27 in groups, and 11 in clusters; 30 exhibit no detectable neighbors. A subset of 67 galaxies occupies extremely low-density regions ($\Sigma_3 D < 0.01$ gal Mpc$^{-3}$), while 22 have their nearest companion farther than 5 Mpc, suggesting their location within extended cosmological voids. The Milky Way ($\Sigma_5 NN \approx 0.13$ gal Mpc$^{-3}$, $R \approx 2.1$ Mpc) and its close analog NGC 3521 both reside in filamentary environments, consistent with intermediate-density surroundings at the boundary of a nearby void. For galaxies with $z > 0.02$, the estimated local densities of the Milky Way and the analyzed systems allow us to identify three additional candidates that satisfy the supplementary environmental density criterion. Based on 3D Voronoi tessellation density estimates, one Milky Way-analog candidate is CGCG 208-043, while according to the fifth-nearest-neighbor approach, the candidates are NGC 5231 and CGCG 047-026. These results highlight the importance of local environmental density as an additional indicator in the search for Milky Way galaxy-analogs.*

***Keywords***: *galaxies, Milky Way, Milky Way galaxy-analogs, voids, Voronoi tessellation.*

## 1. INTRODUCTION

The large-scale structure of the Universe forms a vast network of filaments, walls, clusters, and voids — the cosmic web, that arose as gravity amplified the tiny density fluctuations in the early Universe. This intricate pattern defines the environment in which galaxies form and evolve. It determines how they gather baryonic matter, how frequently they interact or merge, and how much gas remains available to form new stars (e.g., [34]). Galaxies are constantly undergoing evolutionary changes under the influence of their internal physics — such as feedback from supernovae and active galactic nuclei — and by external influences from their surroundings, including tidal forces, ram-pressure stripping, and large-scale flows of matter inside this web. The combined impact of these factors underlies the well-known environmental trends in galaxy properties such as color, morphology, stellar mass, and star formation rate [8]. To understand how the environment governs these processes, it is essential to develop reliable, multiscale approaches to describe the density field and to map the topology of the cosmic web.

Quantifying such an environment, however, requires practical and statistically consistent tools. Over the past decades, several statistical and geometrical methods have been implemented for this purpose, aiming to characterize how galaxies are distributed in space and how their surroundings affect their evolution. The most common approaches — such as the nearest-neighbor and fixed-aperture techniques — provide estimates of local density, though they can be affected by survey geometry, selection effects, and redshift uncertainties. Adaptive methods, such as smoothing kernels and Voronoi tessellation, offer more flexibility and reduce dependence on parameters. An extensive comparison among various estimators of density presented in [8, 62] showed that methods based on Voronoi tessellation provide one of the most efficient and unbiased reconstructions of the density field since they adapt naturally to the intrinsic galaxy distribution without requiring arbitrary smoothing scales. The importance of properly defining the environment density, specifically from the point of view of dark matter halo classification, is shown in [67]: halo properties such as mass, spin, and concentration vary systematically with their position inside filaments, sheets, or voids and thus link the small-scale dynamics of halos to the large-scale topology of the cosmic web.

The Voronoi tessellation method has recently acquired special importance in the extragalactic investigations as a non-parametric, geometrical tool of restoration of the spatial distribution of galaxies. Early applications [41] employed three-dimensional Voronoi tessellation to map the structure of the Local Supercluster, showing that the galaxy distribution forms coherent filaments and walls without any scale assumption a priori. Building on this, the high-order 3D Voronoi technique [20] was elaborated for the detection in redshift space of isolated galaxies, close pairs and triplets, introducing a quantitative criterion of environmental isolation based on the volume ratios of the adjacent cells. These criteria allowed it to perform an automated galaxy classification along the continuous spectrum of environments, from voids up to groups and clusters of galaxies. The Voronoi tessellation and other density-field estimators in the COSMOS survey were used to assess galaxy environment out to $z$~3 [9], demonstrating the method's broad applicability and comparative advantage as well as to the Sloan Digital Sky Survey (SDSS) data for the estimation of galaxy environmental densities [15—17]. These authors demonstrated that this method works in an efficient and stable way within a wide range of redshifts and densities. The Voronoi volume function (VVF) as a cosmological probe of galaxy-environment correlations and large-scale structure was introduced [45], further extending the method's theoretical underpinning. In our works [62, 64], we summarized that the wide applicability of Voronoi tessellations in astronomy covers such important issues as finding cosmic voids and filamentary networks, studies of hierarchical clustering of galaxies, with its high robustness against incompleteness of the survey, and the ability to reproduce the true large-scale structure topological features.

The cosmic web also provides an essential bridge between dark matter halos, the formation of cosmic structures, and the evolution of galaxies. Results from simulations show that halo properties — including their mass, shape, and angular momentum — are influenced not only by the local density but also by where the halos sit within the web. For example, it was found [23] that halos located along filaments are typically more massive and dynamically older than in voids. As a result, the large-scale geometry of the Universe helps to define the initial conditions under which galaxies form and evolve (e.g., [3]). In other words, many of the galaxy's features we observe today can be traced back to the structure of the cosmic web that surrounds them. A proper understanding of galaxy evolution thus needs to take into account both local and large-scale environmental effects within a single framework. As put by [8], galaxies are sculpted by processes that span a continuum of scales, from the sub-megaparsec domains of groups and interactions to the tens-of-megaparsec structures of filaments and voids (e.g., [33]). At this point, one can combine Voronoi-based density fields with cosmic web topology in order to enable physically motivated modeling that actually uncovers these scales and follows the environmental regulation of star formation, morphology, and active galactic nucleus activity. In this sense, the current study explores how galaxies reflect the geometry of the cosmic web and how the environment acts as an important regulator of their evolution, making use of geometric and statistical diagnostics across multiple spatial scales. When considering the morphology of galaxies, the most interest is pointed to the early and late morphological types as well as to the morphological features of galaxies, e.g., with bar, ring and polar ring, dust lane, merging, etc. (see, for example, our works on the machine and deep learning morphological classification of galaxies at $z < 0.1$ [14, 16, 29, 60, 61, 65].

In this article, we will concentrate on the study of the environmental density and distribution of **isolated galaxies with bars** in the Local Volume. Yet early studies of isolated galaxy simulations have shown that the bars are attributes of the isolated evolution of disk galaxies [44, 59], which can form during pre-encounters with other galaxies, be disrupted, and again form [21]. Playing a crucial role in establishment of internal host's galaxy properties (star formation, disk matter halo, angular momentum, bar-bulge stellar populations, "bar — spiral arms" and "bar — interstellar media" interactions, "bar — supermassive black hole" feedback, etc.), they evolve in a complex way (see, also, references on the numerical simulations of bar properties in the host galaxies at low- and high redshifts, in article [2]. The isolated galaxies with bars are of high interest when considering the **Milky Way galaxy-analogs** (MWAs) [10, 13, 19, 30, 42, 63]. The N-body simulations [43] allowed, in particular, to determine both the orbital parameters of merging with two density profile breaks at ~15—18 kpc and 30 kpc as well as the distribution of stellar and dark matter mass between Gaia-Sausage-Enceladus and the Milky Way. The position of the Local Void adjacent to the Local Group [39] and the Milky Way moving away from this void can play a certain role in avoiding the collision with M31. So, the Milky Way bar is a long-lived structure. The Feedback in Realistic Environments (FIRE) simulations of the bar formation in 13 Milky Way-mass galaxies showed [2] that three of the 13 simulated galaxies form bars (mean bar radius ~1.53 kpc) as a result of tidal interactions, and five as a result of internal evolution of the disk, and that interactions with satellites did not have a sufficient influence on this formation.

Our study aims to estimate the environmental density of isolated galaxies with bars as a whole in the Local Volume using various isolation criteria and to confirm the positions of the Milky Way galaxy-analogs in the nearest cosmic web. We describe the sample of candidates to MWAs in Section 2 and the Python code for the estimation of environmental density in Section 3. The results are discussed in Section 4.

## 2. SAMPLES OF THE ISOLATED GALAXIES WITH BAR AND MILKY WAY GALAXY-ANALOGS

The 2MASS Isolated Galaxies (2MIG) catalog [27] is the basis for our investigation. It consists of 3227 galaxies selected as isolated according to spatial and photometric criteria. A galaxy is regarded as isolated in a case when it has no neighboring systems of comparable brightness, within one magnitude in the $K$ band, within a projected radius of 500 kpc, and with a radial velocity difference smaller than 500 km/s. This strict criterion minimizes environmental influence and diminishes the possibility of recent gravitational interactions that could significantly impact internal galactic dynamics or AGN triggering.

Morphological types of galaxies in the 2MIG catalog are given by the de Vaucouleurs T-type system, originally defined in [11] and modified in the Third Reference Catalogue of Bright Galaxies (RC3) [12]. In the present work, we adopt the extended RC3 morphological scale as implemented in the HyperLEDA database [36], which provides homogeneity and comparable morphological comparisons for a wide range of galaxy populations. Based on the Milky Way-analog selection criteria proposed in [63], we focus on barred galaxies of types SB, SAB, and SB0, obtaining a subset of 760 objects. We additionally incorporated 24 MWAs candidates from [49, 50], bringing the total number to 784 galaxies. Following cross-matching with SDSS DR17, we obtained a final dataset of 311 galaxies, which serves as the basis for our subsequent analysis. In this way, we considered only MWAs candidates from the Northern hemisphere, where the most galaxies are located in the redshift range 0.01—0.05 (Figure 1).

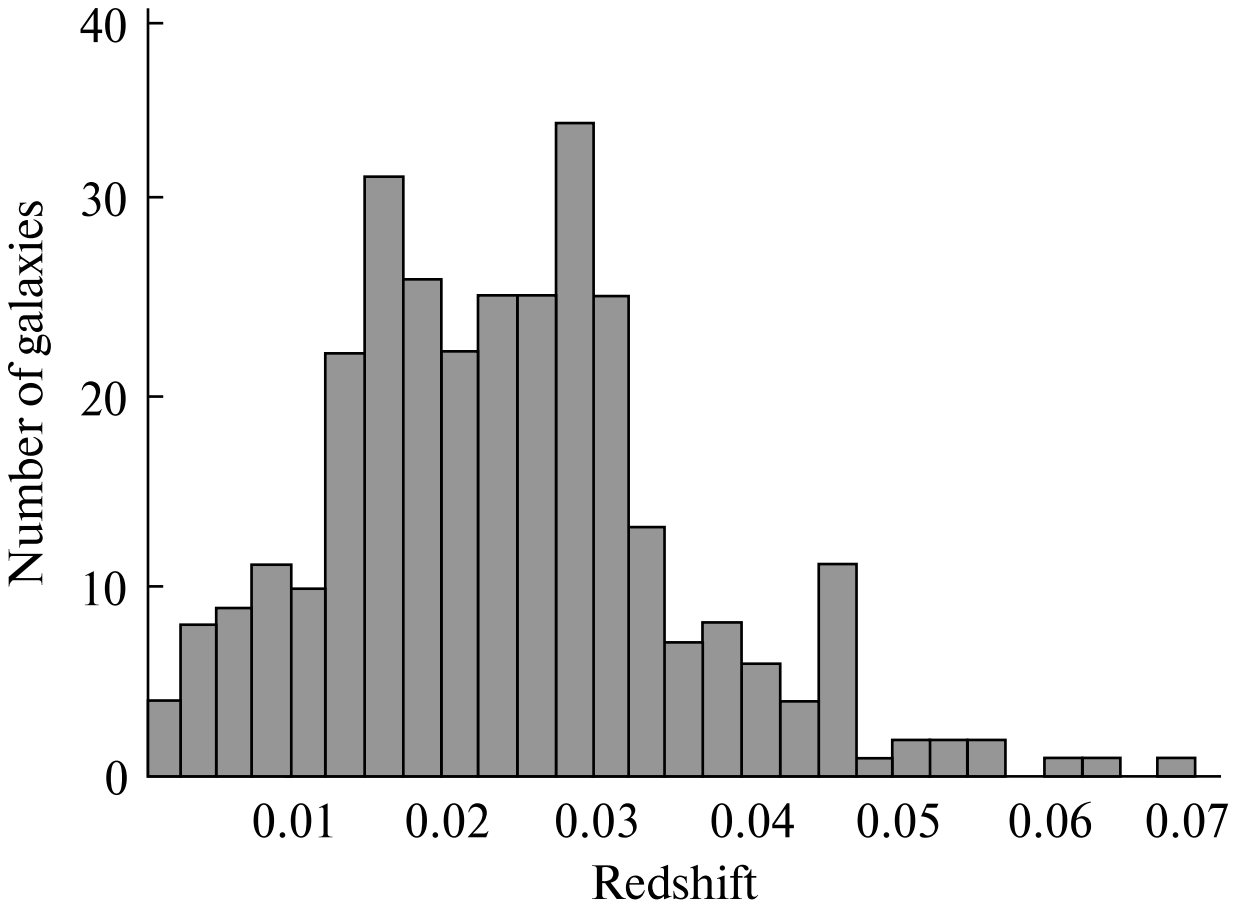

***Figure 1.*** Distribution of the sample of MWAs candidates by redshift (taken from 2 MIG catalog and crossmatched with SDSS DR17)

## 3. THE PYTHON CODE FOR LOCAL ENVIRONMENT ESTIMATION FOR THE GALAXIES WITH BARS AND MILKY WAY GALAXY-ANALOG CANDIDATES

There are several well-established methods to measure properties of the galaxy environment. For each target galaxy, we have equatorial coordinates (*RA, Dec*) and spectroscopic redshift $z_0$ using the SDSS DR17 database, selecting the spectroscopic neighbours through SQL commands. We choose a physical search radius of $R_{phys} = 10$ Mpc, angular radius of the search area given by the angular-diameter distance $D_{A(z0)}$. Only primary, clean galaxy spectra are used for training and validation (sciencePrimary=1 OR legacyPrimary=1; clean=1; type=3; zWarning=0), classifying objects as GALAXY, QSO, or AGN.

**3.1.** ***Low- and high z-regimes.*** Since peculiar motions dominate at very low redshift, we adopt two different regimes for the line-of-sight selection:

• **Low-z regime**: $z_0$ **< 0.02.** A rest-frame velocity constraint $|\Delta v| \le 500$ km/s. (softened to 700 km/s if the number of candidates is too low).

• **High-z regime**: $z_0 \ge$ **0.02**

$$\Delta z = H(z_0) \times (1 + z_0) \times R_{phys}/c, \qquad (1)$$

with $H(z_0)$ calculated with Planck18 cosmology (Astropy).

We explicitly discard after SQL cone selection:

• internal features of the target (its SDSS parentID);

• objects within max(30″, 3 × petro$R_{90}$) from its center;

• very close duplicates (≤1″);

• inhomogeneous target assignment (e.g., duplicate entries having the same parentID, retaining the largest petro$R_{90}$).

**3.2.** ***The 3D coordinate transformation.*** Astrometric positions are transformed into 3D comoving Cartesian coordinates, relative to the target galaxy. Distances are converted to physical distances, dividing by $(1 + z_0)$. For each neighbor with angular position $(RA, Dec)$ :

$$x = D_{phys} \times \cos(Dec) \times \cos(RA),$$
$$y = D_{phys} \times \cos(Dec) \times \sin(RA), \quad (2)$$
$$z = D_{phys} \times \sin(Dec),$$

where $D_{phys} = D_{com}/(1 + z_0)$. This brings the target to an origin position.

At $z_0 < 0.02$, the LOS distances are not trustworthy and only the projected physical separations are used: $R_p = D_A(z_0)$.

**3.3.** ***Environment density estimators.*** We use several different complementary proxies for the local galaxy density, which we select depending on the redshift range and available neighbour statistics.

**3.3.1.** *The 3D estimators ($z_0 \geq 0.02$).* ***k*-nearest neighbours** (kNN) in 3D physical space: for the $k$-th neighbour at radius $R_k$,

$$\Sigma_{\rm simple} = k \times (\pi\, 4/3 \times R_k^{\,3})^{-1}, \quad (3)$$

where $k \in \{3, 5\}$. This gives a good backup when neighbour numbers are not too high.

**3D Voronoi tessellation** using **scipy.spatial.Voronoi**: for a bounded target cell with volume $V_{cell}$,

$$\Sigma_{\rm 3D_Vor} = 1 \times V_{cell}^{\;-1} . \quad (4)$$

Cells that are unbounded, degenerate, on the global hull, or intersect the 10-Mpc search boundary (tested with a safety margin of 0.2 Mpc) are flagged and excluded from the Voronoi density. For reliable 3D tessellation, we require at least 10 neighbours within $R_{phys}$.

**3.3.2.** *The 2D estimators* ($z_0 < 0.02$). Two-dimensional analysis considers solely projected physical separations, hence no line-of-sight details. This methodology is appropriate at low redshifts since large peculiar velocities are observed. The large arbitrariness in 3D distribution simply does not allow for an accurate restoration. The Voronoi-based method is the preferred estimator for surface density:

$$\Sigma_{\rm 2D_Vor} = 1 \times A_{cell}^{\;-1}.$$

For numerous galaxies, the surface density requires orientation without a sufficiently closed Voronoi cell. We achieve that using the $k$-nearest-neighbour approach:

$$\Sigma_{\rm kNN,2D} = k\,(\pi\, 4/3 \times R_p^{\,2})^{-1},$$

where $k \in \{3,5\}$. Our final 2D surface density mostly relies on $\Sigma_{\rm 2D_Vor}$. When it cannot be calculated, we implement this parameter's surrogate, $\Sigma_{\rm kNN,2D}$. The analogous procedure is applied for three-dimensional analysis, although the Voronoi-based density estimator $\Sigma_{\rm 3D_Vor}$ is preferable, since the loci of the local density field are a better fit. When a galaxy's associated Voronoi cell is unbounded due to insufficient nearby points, however, $\Sigma_{\rm kNN5,3D}$ is used. If the number of neighbors within our search radius is less than five, the fallback estimator $\Sigma_{\rm kNN3,2D}$ is instead applied. This approach was already verified for Polar Ring Galaxies (see [17]. The final environmental label is assigned to the best available estimator (preferably $\Sigma_{\rm 3D_Vor}$, fallback to $\Sigma_{simple}$) according to empirical thresholds from [51, 52] given in Table 1.

Additionally, this approach allows us to verify the isolation of galaxies in our sample, which includes isolated galaxies with bars as well as Milky Way analogs, as described in Section 1. We consider a galaxy to be isolated if it has no neighbour within a projected distance of 500 kpc and resides in a low-density environment, corresponding to the *void* or *filament* type according to the classification criteria listed in Table 1.

**3.4.** ***Implementation and deliverables.*** Our Python-based pipeline employs a combination of libraries including *NumPy*, *pandas*, *Astropy*, and astroquery.sdss for data handling, *SciPy.spatial* and *Shapely* for geometric analyses, and *Matplotlib* together with *PyVista* for visualization and interactive 3D HTML export. For each galaxy, the output includes neighbour statistics, k-nearest-neighbour (kNN) radii and densities, Voronoi-based density measures, diagnostic flags, and an environmental label.

*Table 1.* **Environmental density classification of galaxies**

| Environment | Criterion (Number of galaxies × $Mpc^{-3}$) |
|---|---|
| Void | $\Sigma < 0.05$ |
| Filament | $0.05 \leq \Sigma < 0.5$ |
| Group | $0.5 \leq \Sigma < 2.0$ |
| Cluster | $\Sigma \geq 2.0$ |

## 4. RESULTS AND DISCUSSION

As a result of implementation of the Python code for local density environment estimation for the galaxies with bars and the Milky Way galaxy-analog candidates, 157 galaxies were classified as galaxies in a void, 84 galaxies as galaxies in a filament, 27 as galaxies in a group, 11 as galaxies in a cluster; for 30 galaxies, this code did not find any neighbours (Figure 2). The absence of detected neighbors is most often associated either with reliable SDSS spectroscopic data that are unavailable or with the galaxy being located near the edge of the SDSS survey footprint, where neighbor information is incomplete. In Figure 3, we show examples of 2D and 3D environmental density estimates based on Voronoi tessellation for the studied galaxies. While 3D Voronoi tessellation enables us to reconstruct the large-scale structure of the Universe — which is especially important if considering large samples of galaxies — the 2D projection of the 3D galaxy distribution results in some level of overestimation of the local galaxy surface density. Therefore, in the case of galaxies located at redshifts $z < 0.02$, we expect that the true environmental densities differ by several times from the projected ones. Unlike 2D Voronoi cells, 3D Voronoi tessellation provides a much more accurate estimate of local density, which is consistent with results obtained by other methods. Moreover, this approach makes it possible to reproduce the shape of the Voronoi cell of the central galaxy in 3D space, allowing for a more realistic visualization of its galaxy neighborhood.

A total of 67 galaxies are located in extremely low-density environments, with local 3D densities below 0.01 gal Mpc$^{-3}$, and 19 galaxies reside in regions of extremely low projected surface density, with values below 0.01 gal Mpc$^{-2}$. For most of these systems, it was not possible to construct bounded 3D or 2D Voronoi cells; therefore, the local density was estimated based on the distance to the first, third, or fifth nearest neighbour. Among these galaxies at $z > 0.02$, one system has its nearest neighbour within 1 Mpc, 11 within the range 1-2 Mpc, 30 between 2—5 Mpc, and the remaining 22 have their closest companion at distances greater than 5 Mpc. Naturally, these results require further verification using independent methods. However, the preliminary analysis suggests that the 22 galaxies located more than 5 Mpc away from their nearest neighbors are likely situated within large cosmological voids.

We verified whether the galaxies classified in our sample belong to cosmological voids using the catalog [37], which was constructed with the *VIDE void*

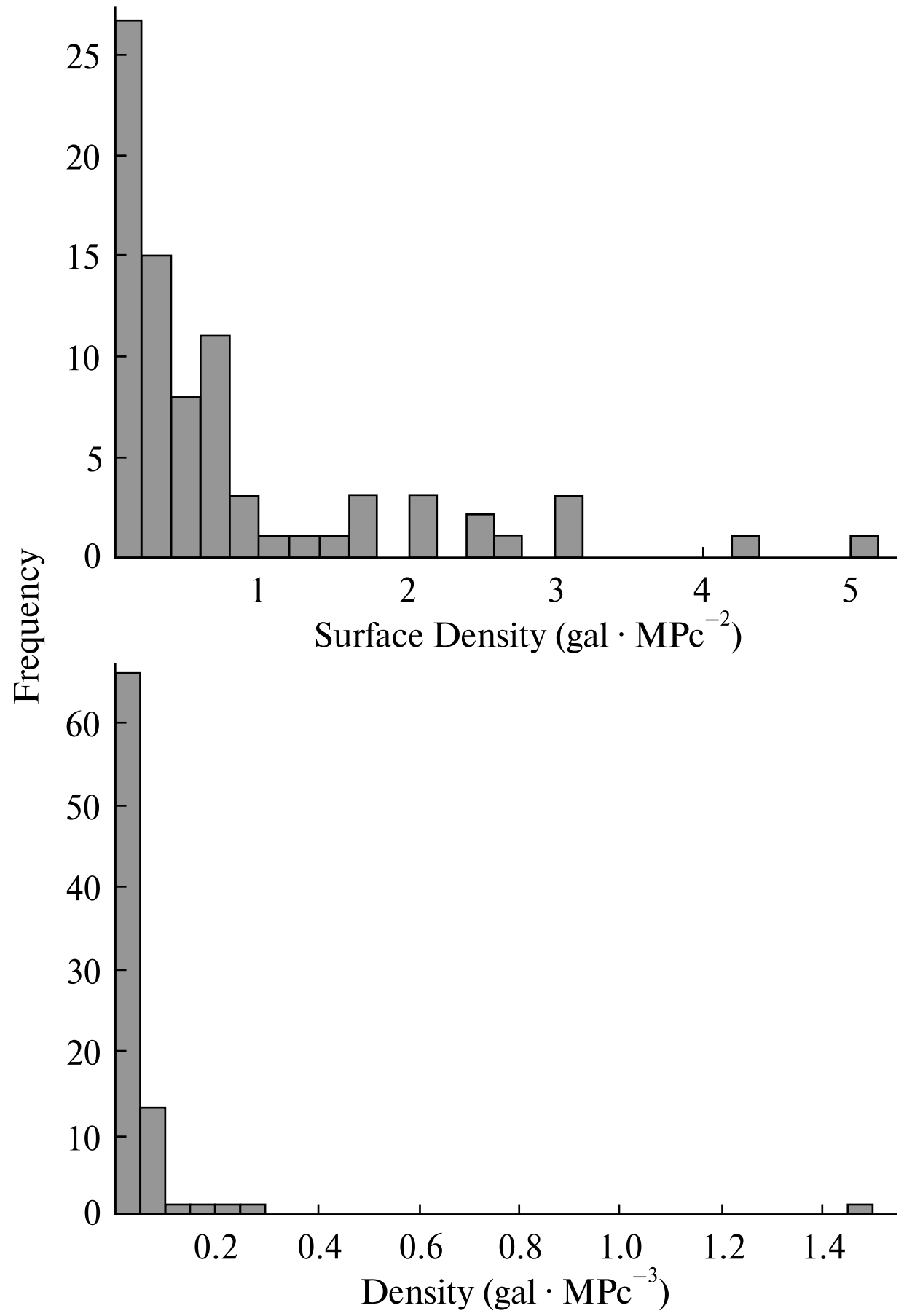

***Figure 2.*** The $\Sigma_{2D_Vor}$ distribution for galaxies in the sample at $z < 0.02$ (top panel) and the $\Sigma_{3D_Vor}$ distribution for galaxies in the sample at $z \geq 0.02$ (bottom)

*finder* approach. For each galaxy and void center, we calculated the comoving distances $D_{c(z)}$ and their corresponding Cartesian coordinates $(x, y, z)$. It allows us to work with three-dimensional positions of objects (in Mpc) rather than their projection on the celestial sphere. In the catalog mentioned, each void is approximated by a sphere with a given radius $R_{void}$ and a center defined at the "redshift center". Galaxy-void association was done using a KD-tree algorithm. With this method, for each galaxy, we efficiently searched for the nearest void center. The distance between a galaxy and a void center was computed as Euclidean in comoving coordinates. A galaxy is deemed to lie within a given void if its distance to the center satisfies $d \leq R_{void}$, and if its redshift $z$ lies within the range from $z_{near}$ to $z_{far}$, accounting for a possible velocity uncertainty of ≈ 300 km/s along the line of sight. In summary, out of 311 galaxies with bars in our sample, 113 can be identified as lying inside the interior of the voids listed in the catalog. Among them, 16 galaxies lack a sufficient number of neighbors to construct a "crow's nest" around the central galaxy according to our approach. The remaining systems have their nearest neighbors located at distances greater than 1 Mpc, with only a few galaxies having closer companions. These results suggest that nearly one-third of the sample is situated in low-density regions corresponding to cosmological or local voids.

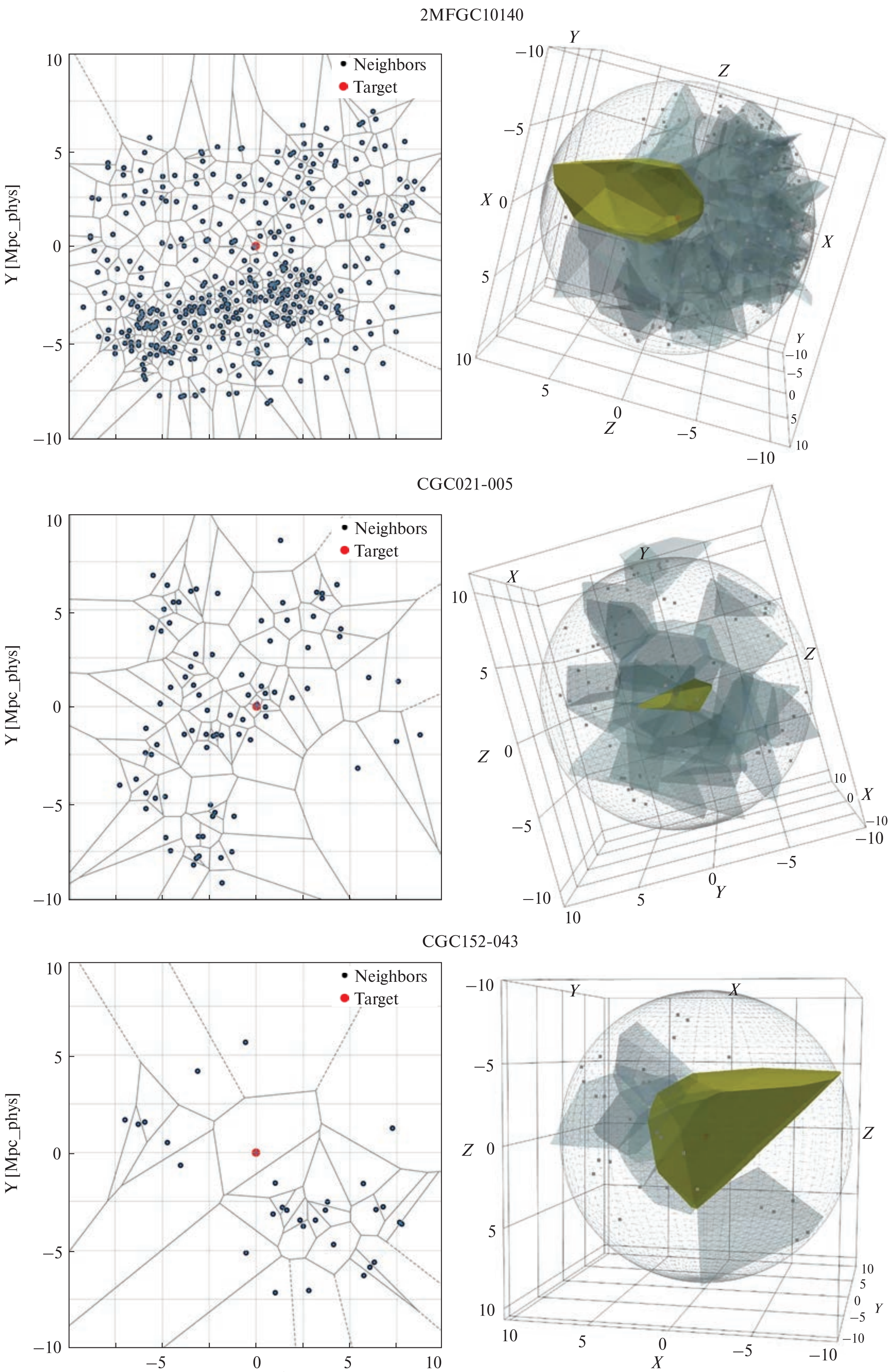

*Figure 3.* The 2D (left panel) and 3D (right panel) Voronoi tessellation for MWAs candidates

Thus, having the environmental density estimates of galaxies, we proceed to analyze their ionization properties using optical emission-line diagnostics. We used the Baldwin—Phillips—Terlevich (BPT, [5]) diagram to determine the main sources of ionization and distinguish star-forming ones from composite systems and galaxies with active galactic nuclei (AGN). The AstroPy library allows us to execute individual SQL queries by sky coordinates with a 0.01° search radius. We found spectra for 147 out of 309 galaxies with bars. The fluxes of the Hα, Hβ, [N II] λ 6584, and [O III] 5007 emission lines are the standard input variables to the BPT diagnostic with the classification curves of other diagnostics by [26, 28, 57] — star-forming, composite, LINER, and Seyfert types (Figure 4). So, we get 48 star-forming systems, 41 composite galaxies, 45 LINERs, 12 Seyferts, and 2 unclassified objects among the 147 isolated galaxies with bar, having spectroscopic data.

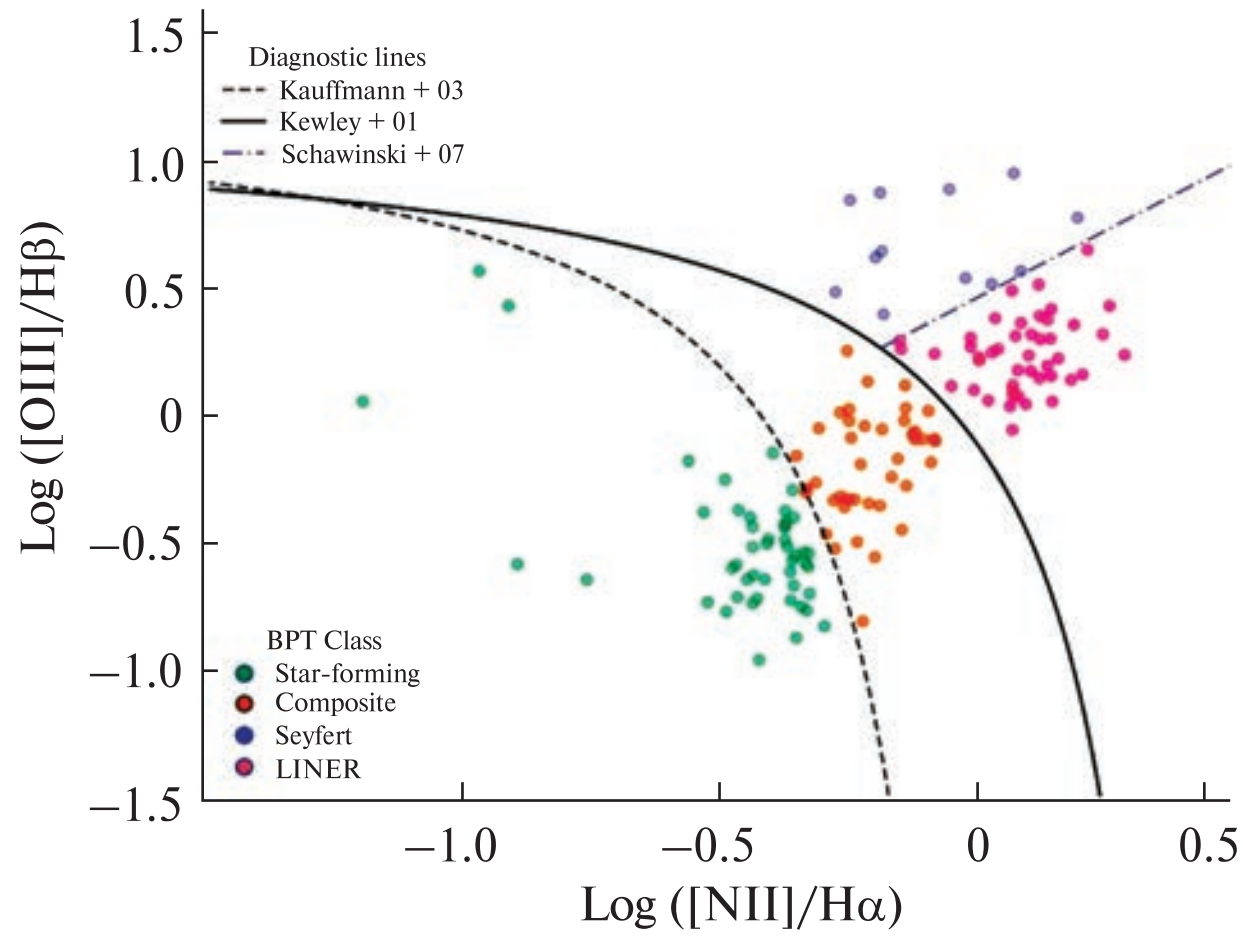

***Figure 4.*** The BPT diagram for galaxies with bars selected from the 2MIG catalog and cross-matched with SDSS DR17

To assess whether the nuclear activity of galaxies traced by the BPT classification correlates with the neighborhood environment, we compared the likelihood distributions of environmental density across the four spectral classes using the Kruskal — Wallis H-test. For the nearby galaxies ($z < 0.02$), we adopted the projected surface density $\Sigma_{2D_Vor}$ to account for seeing effects and projection uncertainties, while for galaxies at higher redshift ($z \geq 0.02$), we used the 3D Voronoi-based density parameter $\Sigma_{3D_Vor}$, or the simplified k-nearest-neighbour estimate ($\Sigma_{simple}$), when Voronoi-based densities were unavailable to apply. The test results are as follows: $H = 2.13$, $p = 0.55$ for $z < 0.02$ and $H = 4.07$, $p = 0.25$ for $z \geq 0.02$. This indicates that the environment density does not significantly differentiate galaxies hosting LINER or Seyfert nuclei from purely star-forming systems. Although the limited sample size (e.g., only three Seyferts at $z < 0.02$) may partly reduce statistical significance, the obtained parameters suggest only a weak or absent influence of the environmental galaxy neighborhood on the activity type of galaxies. These findings imply that AGN triggering in galaxies with bars, as well as Milky Way galaxy-analogs, is not primarily driven by local galaxy density within this redshift range, but rather by internal secular processes or bar-driven gas inflow. This result is consistent with recent numerical simulations and conclusions by [2] that environmental density has no sufficient influence on the bar's formation.

Figure 5 illustrates the boxplots of the logarithmic environmental densities for star-forming, composite, LINER, and Seyfert galaxies. For the nearby galaxies ($z < 0.02$), the projected surface densities suggest that star-forming systems may occupy somewhat lower-density environments, whereas AGN-hosting galaxies (especially Seyfert galaxies) appear in slightly denser regions. However, these apparent trends are not statistically significant, as confirmed by the Kruskal — Wallis H-test. In the higher-redshift regime ($z \geq 0.02$), based on the 3D Voronoi or k-nearest-neighbour estimates, all spectral types span comparable ranges of local environmental density, confirming the absence of measurable segregation by nuclear activity type. The horizontal blue lines in each box denote the median values for each class. This remains broadly consistent across both redshift bins and activity classes, supporting the conclusion of negligible environmental density dependence.

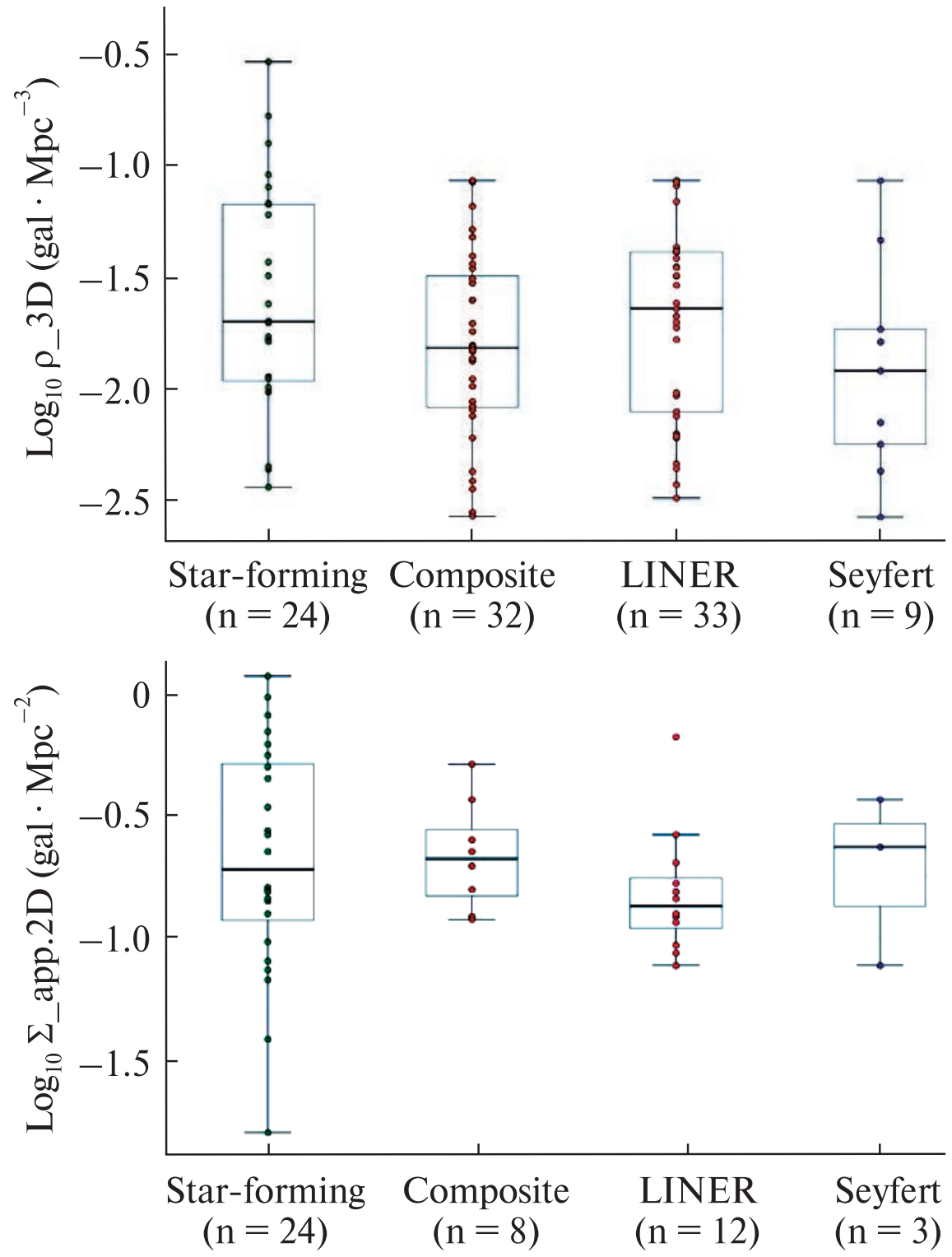

***Figure 5.*** Distribution of 3D (top; $z \geq 0.02$) and local projected (bottom; $z < 0.02$) galaxies densities across BPT activity classes. The blue line within each box represents the median value for the activity types

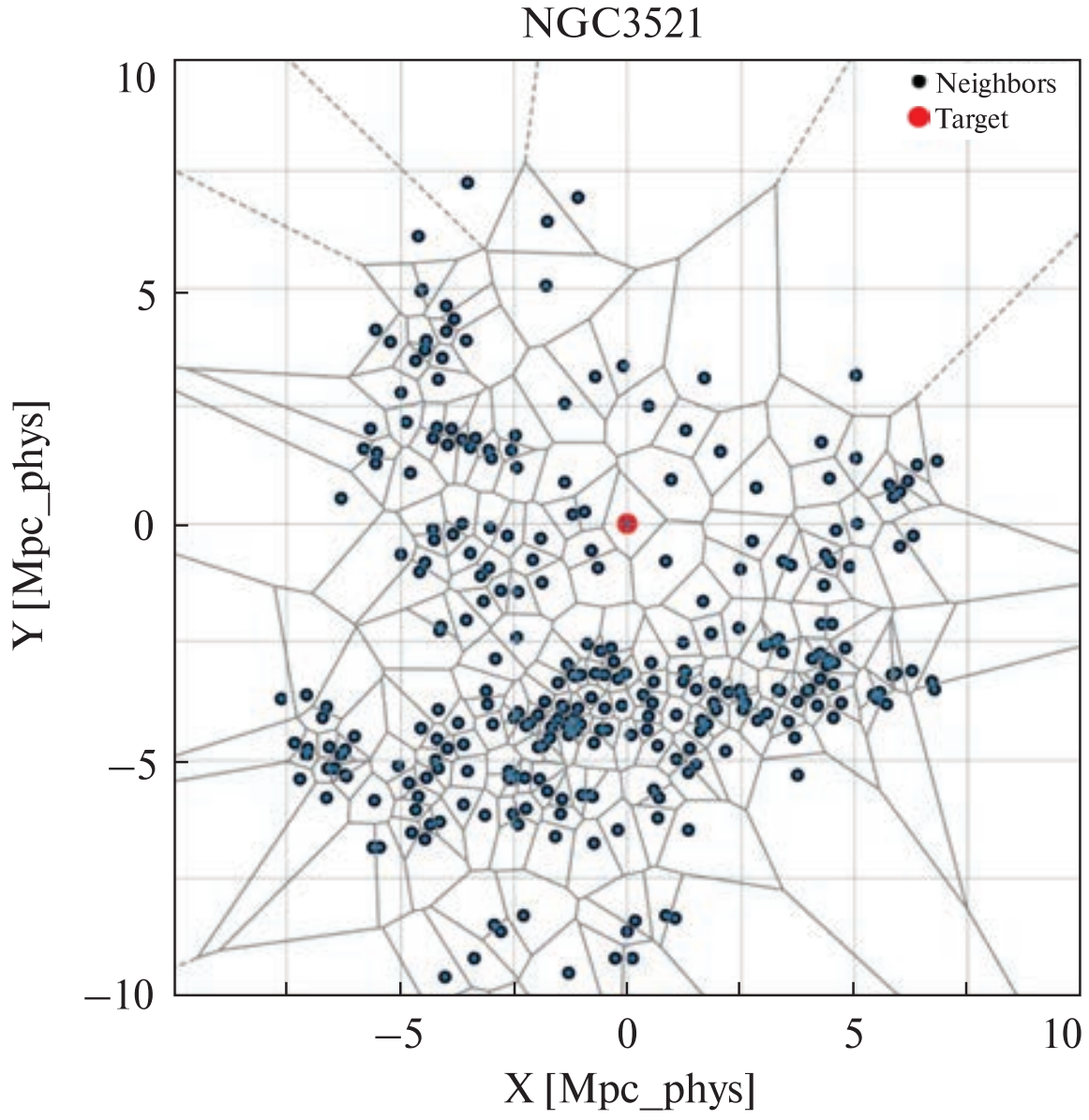

***Figure 6.*** The 2D Voronoi tessellation for Milky Way galaxy-analog NGC 3521

The absence of a statistically significant difference between the types of activity of the studied galaxies with bars indicates that their activity (type) most likely does not depend on the characteristics of the environment neighborhood in the cosmic web. This result is consistent with previous studies showing that activity of AGN and LINER is common even in isolated low-density systems [4, 24, 53, 54, 56, 66]. In this case, the leading processes of fuel delivery are also likely to be internal, including bar-driven gas infall and disk instabilities that provide a channel for active transport to the central region, sustaining mild activity of AGN or circumnuclear star formation [1, 7, 31]. Moreover, as argued in [23], the local environmental density is the key factor influencing the properties of a galaxy's dark matter halo and, consequently, various intrinsic parameters of the galaxy itself. At the same time, the type of the cosmic web structure (filament, wall, or void) has only a minor effect, provided that the local density remains the same. The relatively high fraction of LINERs and composite systems among the studied galaxies with bars and the Milky Way galaxy-analogs confirms that low AGN activity may be a typical evolutionary phase in secular evolution disk galaxies. Similar conclusions have been drawn for isolated or empty galaxies in the Local Universe, where the AGN frequency remains comparable to that in denser regions [32, 55, 56]. These results suggest that the presence of a bar and the presence of internal gas reservoirs are more important factors for triggering and sustaining nuclear activity than the environmental neighborhood density in the cosmic web. This conclusion naturally raises the question of whether similar processes occur in galaxies that are similar in structure and dynamics to the Milky Way. Such MWAs are important reference points for studying how long-term bar-driven evolution and low levels of nuclear activity manifest themselves in disk galaxies with comparable stellar mass and morphology.

The Milky Way-like galaxies, which were modeled within the magneto-hydrodynamic cold-dark-matter simulation IllustrisTNG (TNG50), were analyzed in [48]. These authors derived a set of characteristic properties for these systems, including aspects of their large-scale environment. In particular, they found that no major galaxy lies within 500 kpc of any of the Milky Way analogs. However, the results of the simulation also indicate that most modeled MWAs are located along filaments or sheets of the cosmic web. At $z \sim 0$, none of the MWAs lie in the immediate vicinity of massive clusters, which is consistent with the ~16.5 Mpc distance between the Milky Way and the nearest rich cluster. These authors note that a more rigorous selection would require a dedicated analysis of the isolation criterion for providing such numerical simulations. Therefore, the discrepancy with our findings may arise from differences in how local density parameters are defined in the IllustrisTNG simulation, underscoring the need for a detailed comparison with observationally derived environment neighborhood parameters as presented in this study.

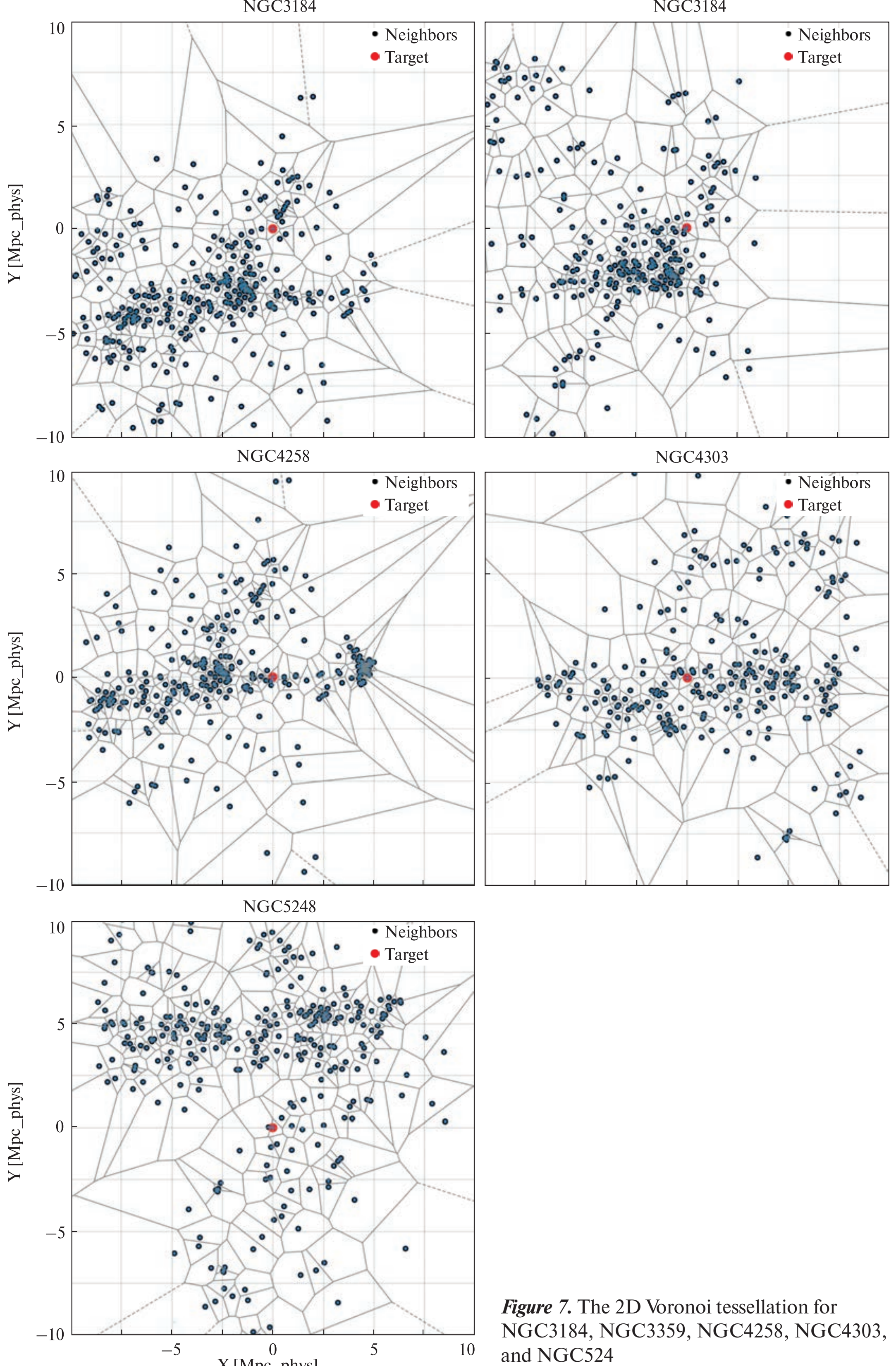

***Figure 7.*** The 2D Voronoi tessellation for NGC3184, NGC3359, NGC4258, NGC4303, and NGC524

According to [40], the Milky Way and the Local Group reside within a sheet-like structure that forms part of the cosmic web, lying at the boundary of the Local Void. This low-to-intermediate density environment is dynamically influenced by the nearby void, which contributes to the peculiar velocity of the Milky Way away from it and slightly toward the Virgo Cluster. Hence, the Milky Way occupies a transitional region between a sheet and a void, rather than a filament or cluster environment.

This general picture is consistent with quantitative estimates of the Milky Way's local environmental density. To evaluate this density, we considered only massive and isolated neighboring galaxies, excluding dwarf companions such as the Magellanic Clouds. The five most significant neighboring systems are Andromeda (M31, 0.77 Mpc, [47]), Triangulum (M33, 0.85 Mpc, [6]), NGC 3109 (≈1.3 Mpc [58]), NGC 300 (≈1.9 Mpc [22]), and NGC 55 (≈2.1 Mpc [25]). Using the distance to the fifth significant neighbor ($R \approx 2.1$ Mpc), the local galaxy density is estimated as $\Sigma_{kNN5} \approx 0.13$ gal Mpc$^{-3}$, following the 5-th nearest neighbor approach, and leads to classify the Milky Way as the galaxy in filament based on our criteria (Section 3) After analyzing the Milky Way's largest neighbors [40], these authors suggested that our Galaxy is located in a wall/sheet. The halos with masses similar to that of the Milky Way situated near the centers of small walls are under-represented according to [23]: they used the Bolshoi—Planck cosmological N-body simulation, which is a ΛCDM model with Planck 2014 parameters. Specifically, these authors analyzed the distribution of dark matter halos in voids, walls, filaments, and nodes. The Milky Way's location in the Local Wall, in the interstitial void between the Local Void and the Sculptor Filament, is indeed statistically rare but entirely consistent with the ΛCDM model. Therefore, we may infer that the local environmental density value could serve as a quantitative descriptor of the surrounding environment, being an additional criterion in the search for Milky Way galaxy-analogs.

One of the best-known analogues of the Milky Way is NGC 3521 [46, 50]. According to our analysis, this galaxy resides in an environment with a projected surface density of approximately 0.7 gal Mpc$^{-2}$, estimated using the Voronoi tessellation method (Figure 6), and 0.4 gal Mpc$^{-2}$ based on the 5-th nearest neighbor approach. Considering possible overestimations due to projection effects, we conclude that NGC 3521 is located in an environment of similar density to that of the Milky Way's surroundings.

Among the eleven additional Milky Way analogs selected by [49, 50], most occupy environments of comparable density, except for NGC 3184, NGC 3359, NGC 4258, NGC 4303, and NGC 5248. These galaxies exhibit projected surface densities exceeding 2 gal Mpc$^{-2}$ (Figure 7) and appear to reside in group- or cluster-like environments, significantly denser than that of the Milky Way or NGC 3521. The derived environmental density parameters require further verification using 3D distances between galaxies in the Local Universe. Nevertheless, these systems do not meet the isolation criterion and should therefore be excluded from the sample of Milky Way analogues.

For galaxies with $z > 0.02$, the estimated local densities of the Milky Way and the analyzed systems allow us to identify three additional candidates that satisfy the supplementary environmental density criterion. Based on 3D Voronoi tessellation density estimates, one Milky Way-analog candidate is CGCG 208-043 (classified as star-forming according to the BPT diagram) , while according to the fifth-nearest-neighbor approach, the candidates are NGC 5231 and CGCG 047-026 (both classified as LINER according to the BPT diagram).

***Acknowledgements.*** *This study is supported by the National Research Fund of Ukraine (Project No. 2023.03/0188, 2024—2026). The work of O. Kompaniiets partially supported by research works of young scientists of the NAS of Ukraine for 2025-2026 (Registration number 0125U002943).*

REFERENCES

1. Alonso S., Mesa V., Padilla N., Lambas D. G. (2012). Galaxy interactions. II. High density environments. *Astron. and Astrophys.*, **539,** id. A46, 9. https://doi.org/10.1051/0004-6361/201117901
2. Ansar S., Pearson S., Sanderson R. E., et al. (2025). Bar Formation and Destruction in the FIRE-2 Simulations. *Astrophys. J.*, **978**, no. 1, Art. no. 37. https://doi.org/10.3847/1538-4357/ad8b45
3. Aragón-Calvo M. A., Platen E., van de Weygaert R. (2010). The Spine of the Cosmic Web. *Astrophys. J.*, **723**, Is. 1, 364—382. https://doi.org/10.1088/0004-637X/723/1/364
4. Argudo-Fernández M., Shen S., Sabater J., et al. (2016). The effect of local and large-scale environments on nuclear activity and star formation. *Astron. and Astrophys.*, **592**, 13. https://doi.org/10.1051/0004-6361/201628232
5. Baldwin J. A., Phillips M. M., Terlevich R. (1981). Classification parameters for the emission-line spectra of extragalactic objects. *Publ. Astron. Soc. Pacif.*, **93**, 5—19. https://doi.org/10.1086/130766
6. Bono G., Caputo F., Marconi M., Musella I. (2010). Insights into the Cepheid Distance Scale. *Astrophys. J.*, **715**(1), 277—291. https://doi.org/10.1088/0004-637X/715/1/277
7. Cheung E., Athanassoula E., Masters K. L., et al. (2013). Galaxy Zoo: Observing Secular Evolution through Bars. *Astrophys. J.*, **779**, Is. 2, id.A162, 18. https://doi.org/10.1088/0004-637X/779/2/162
8. Cybulski R. (2016). *The Cosmic Web and the Role of Environment in Galaxy Evolution*. PhD Thesis, University of Massachusetts.
9. Darvish B., Mobasher B., Sobral D., et al. (2015). A Comparative Study of Density Field Estimation for Galaxies. *Astrophys. J.*, **805**, Is. 2, id. 121, 1—19. https://doi.org/10.1088/0004-637X/805/2/121
10. Denyshchenko S. I., Fedorov P. N., Akhmetov V. S., et al. (2024). Determining the parameters of the spiral arms of the Galaxy from kinematic tracers based on Gaia DR3 data. *Mon. Notic. Roy. Astron. Soc.,* **527**, Is. 1, 1472—1480. https://doi.org/10.1093/mnras/stad3350
11. de Vaucouleurs G. (1959). *Classification and Morphology of External Galaxies*. Handbuch der Physik, **53**, 275 p. https://doi.org/10.1007/978-3-642-45932-0_7
12. de Vaucouleurs G., de Vaucouleurs A., Corwin H. G., Jr., Buta R. J., Paturel G., Fouqu  P. (1991). *Third Reference Catalogue of Bright Galaxies*. New York: Springer, 632 p. ISBN 0-387-97552-7
13. Dmytrenko A. M., Fedorov P. N., Akhmetov V. S., et al. (2025). Spatial orientation and shape of the velocity ellipsoids of the Gaia DR3 giants and sub-giants in the Galactic plane. *Mon. Notic. Roy. Astron. Soc.,* **542**, Is. 3, 2542—2559. https://doi.org/10.1093/mnras/staf1408
14. Dobrycheva D. V., Hetmantsev O. O., Vavilova I. B., et al. (2025). Discovery of the Polar Ring Galaxies with deep learning. *Astron. and Astrophys.*, **702**, A258, 1—13. https://doi.org/10.1051/0004-6361/202555052
15. Dobrycheva D., Melnyk O., Elyiv A., et al. (2016). Environmental density of galaxies from SDSS via Voronoi tessellation. *Proc. IAU*, **308**, 248—249. https://doi.org/10.1017/S1743921316009959
16. Dobrycheva D. V., Melnyk O. V., Vavilova I. B., et al. (2014). Environmental Properties of Galaxies at $z < 0.1$ from the SDSS via the Voronoi Tessellation. *Odessa Astron. Publ.*, **27**, 26—27.
17. Dobrycheva D. V., Melnyk O. V., Vavilova I. B., et al. (2015). Environmental Density vs. Colour Indices of the Low Redshifts Galaxies. *Astrophysics*, **58**, Is. 2, 168—180. https://doi.org/10.1007/s10511-015-9373-x
18. Dobrycheva D. V., Vavilova I. B., Khramtsov V. P., et al. (2025). The visual vs. CNN verification of catalogs of SDSS merging galaxies, galaxies with bar, ring, and dust lane. *Astron. and Astrophys*.
19. Dobrycheva D., Vavilova I., Khramtsov V., et al. (2025). Machine Learning Mismatchings and Catalogues Creation: A Path to Finding the Milky Way Galaxies-Analogues. *ASP Conf. Ser.*, ISSN (ISSN-L) 1050-3390.
20. Elyiv A., Melnyk O., Vavilova I. B. (2008). High-order 3D Voronoi tessellation for identifying Isolated galaxies, Pairs and Triplets. *Mon. Notic. Roy. Astron. Soc.,* **394**, Is. 3, 1409—1418. https://doi.org/10.1111/j.1365-2966.2008.14150.x
21. Ghosh S., Di Matteo P. (2024). Looking for a needle in a haystack: Measuring the length of a stellar bar. *Astron. and Astrophys.*, **683**, Art. no. A100. https://doi.org/10.1051/0004-6361/202347763
22. Gieren W., Pietrzyński G., Soszyński I., et al. (2005). The Araucaria Project: Near-Infrared Photometry of Cepheid Variables in the Sculptor Galaxy NGC 300. *Astrophys. J.*, **628**(2), 695—703. https://doi.org/10.1086/430903
23. Goh T., Primack J., Lee C. T., et al. (2019). Dark matter halo properties versus local density and cosmic web location. *Mon. Notic. Roy. Astron. Soc.,* **483**, Is. 2, 2101—2122. https://doi.org/10.1093/mnras/sty3153

24. Hernández-Ibarra F. J., Dultzin D., Krongold Y., et al. (2013). Nuclear activity in isolated galaxies. *Mon. Notic. Roy. Astron. Soc.,* **434**, Is. 1, 336—346. https://doi.org/10.1093/mnras/stt1021
25. Jacobs B. A., Rizzi L., Tully R. B., et al.(2009). The Extragalactic Distance Database: Color—Magnitude Diagrams. *Astrophys. J.*, **138**(2), 332—337. https://doi.org/10.1088/0004-6256/138/2/332
26. Kauffmann G., Heckman T. M., Tremonti C., et al. (2003). The host galaxies of active galactic nuclei. *Mon. Notic. Roy. Astron. Soc.,* **346**, Is. 4, 1055—1077. https://doi.org/10.1111/j.1365-2966.2003.07154.x
27. Karachentseva V. E., Mitronova S. N., Melnyk O. V., et al. (2010). Catalog of Isolated Galaxies Selected from the 2MASS Survey. *Astrophys. Bull.*, **65**, Is. 1, 1—17. https://doi.org/10.1134/S1990341310010013
28. Kewley L. J., Dopita M. A., Sutherland R. S., et al. (2001). Theoretical Modeling of Starburst Galaxies. *Astrophys. J.*, **556**, Is. 1, 121—140. https://doi.org/10.1086/321545
29. Khramtsov V., Vavilova I. B., Dobrycheva D. V., et al. (2022). Machine learning technique for morphological classification of galaxies from the SDSS. III. Image-based inference of detailed features. *Space Sci. and Technol.*, **28**, Is. 5, 27—55. https://doi.org/10.15407/knit2022.05.027
30. Khramtsov V., Dobrycheva D., Vavilova I., et al. (2025). Vision-Language Models for Spiral Galaxy Identification in SDSS: A Path to Finding Milky Way Analog Galaxies. *ASP Conf. Ser.*, ISSN (ISSN-L) 1050-3390.
31. Kim T., Gadotti D. A., Athanassoula E., et al. (2016). Evidence of bar-induced secular evolution in the inner regions of stellar discs in galaxies: what shapes disc galaxies? *Mon. Notic. Roy. Astron. Soc.,* **462**, Is. 4, 3430—3440. https://doi.org/10.1093/mnras/stw1899
32. Kuutma T., Tamm A., Tempel E. (2017). From voids to filaments: environmental transformations of galaxies in the SDSS. *Astron. and Astrophys.*, **600**, id. L6, 5. https://doi.org/10.1051/0004-6361/201730526
33. Laigle C., Pichon C., Arnouts S., et al. (2018). COSMOS2015 photometric redshifts probe the impact of filaments on galaxy properties. *Mon. Notic. Roy. Astron. Soc.,* **474**, Is. 4, 5437—5458. https://doi.org/10.1093/mnras/stx3055
34. Libeskind N. I., van de Weygaert R., Cautun M., et al. (2018). Tracing the Cosmic Web. *Mon. Notic. Roy. Astron. Soc.,* **473**, Is. 1, 1195—1217. https://doi.org/10.1093/mnras/stx1976
35. Lindner U., Einasto J., Einasto M., et al. (1995). The structure of supervoids. I. Void hierarchy in the Northern Local Supervoid. *Astron. and Astrophys.*, **301**, 329. https://doi.org/10.48550/arXiv.astro-ph/9503044
36. Makarov D., Prugniel P., Terekhova N., et al. (2014). HyperLEDA. III. The catalogue of extragalactic distances. *Astron. and Astrophys.*, **570**, id. A13, 1—12. https://doi.org/10.1051/0004-6361/201423496
37. Malandrino R., Lavaux G., Wandelt B. D., et al. (2025). A Bayesian catalog of 100 high-significance voids in the Local Universe. ArXiv:2507.06866. https://doi.org/10.48550/arXiv.2507.06866
38. Marius C., van de Weygaert R., Bernard J. T. J., et al. (2014) Evolution of the cosmic web. *Mon. Notic. Roy. Astron. Soc.,* **441**, Is. 4, 2923—2973. https://doi.org/10.1093/mnras/stu768
39. Mazurenko S., Banik I., Kroupa P., et al. (2024). A simultaneous solution to the Hubble tension and observed bulk flow within 250 $h_{-1}$ Mpc. *Mon. Notic. Roy. Astron. Soc.*, **527**, Is. 3, 4388—4396. https://doi.org/10.1093/mnras/stad3357
40. McCall M. L. (2014). A Council of Giants. *Mon. Notic. Roy. Astron. Soc.,* **440**, Is. 1, 405—426. https://doi.org/10.1093/mnras/stu199
41. Melnyk O., Elyiv A. A., Vavilova I. B. (2007). The Structure of the Local Supercluster by 3D Voronoi Tessellation. *Kinemat. Phys. Celest. Bodies*, **22**, Is. 4, 283—296. arXiv:0712.1297 https://www.mao.kiev.ua/index.php/ua/pdf-opener?vavilova/Melnyk-Elyiv-Vavilova-KPCB-2006-0712.1297
42. Mishenina T., Gorbaneva T., Dmytrenko A., et al. (2024). Specific Features of the Enrichment of Metal-Poor Stars with Neutron-Capture R-Process Elements. *Odessa Astron. Publ.,* **37**, 47—51. https://doi.org/10.18524/1810-4215.2024.37.312691
43. Naidu R. P., Conroy C., Bonaca A., et al. (2021). Reconstructing the Last Major Merger of the Milky Way with the H3 survey. *Astrophys. J.*, **923**, Is. 1, art. id. 92, 24 p. https://doi.org/10.3847/1538-4357/ac2d2d
44. Ostriker J. P., Peebles P. J. E. (1973). A Numerical Study of the Stability of Flattened Galaxies: or, can Cold Galaxies Survive? *Astrophys. J.*, **186**, 467—480. https://doi.org/10.1086/152513
45. Paranjape A., Alam S. (2020). Voronoi volume function: a new probe of cosmology and galaxy evolution. *Mon. Notic. Roy. Astron. Soc.,* **495**, Is. 3, 3233—3251. https://doi.org/10.1093/mnras/staa1379
46. Pastoven O. S., Kompaniiets O. V., Vavilova I. B., Izviekova I. O. (2024). NGC 3521 as the Milky Way analogue: Spectral energy distribution from UV to radio and photometric variability. *Space Sci. and Technol.*, **30**, Is. 6(151), 67—83. https://doi.org/10.15407/knit2024.06.067
47. Paturel G., Teerikorpi P., Theureau G., et al. (2002). Calibration of the distance scale from galactic Cepheids. II. Use of the HIPPARCOS calibration. *Astron. and Astrophys.*, **389**, 19—28. https://doi.org/10.1051/0004-6361:20020492
48. Pillepich A., Sotillo-Ramos D., Ramesh R. (2024). Milky Way and Andromeda analogues from the TNG50 simulation. *Mon. Notic. Roy. Astron. Soc.,* **535**, Is. 2, 1721—1762. https://doi.org/10.1093/mnras/stae2165

49. Pilyugin L. S., Grebel E. K., Kniazev A. Y. (2014). The Abundance Properties of Nearby Late-type Galaxies. I. The Data. *Astrophys. J.*, **147**, Is. 6, id. 131, 1—24. https://doi.org/10.1088/0004-6256/147/6/131
50. Pilyugin L. S.; Tautvaišienė G.; Lara-López M. A. (2023) Searching for Milky Way twins: Radial abundance distribution as a strict criterion. *Astron. and Astrophys.*, **676**, id. A57, 1—28. https://doi.org/10.1051/0004-6361/202346503
51. Poggianti B. M., Desai V., Finn R. (2008). The Relation between Star Formation, Morphology, and Local Density in High-Redshift Clusters and Groups. *Astrophys. J.*, **684**, Is. 2, 888—904. https://doi.org/10.1086/589936
52. Poggianti B. M.; De Lucia G.; Varela J. (2010). The evolution of the density of galaxy clusters and groups: denser environments at higher redshifts. *Mon. Notic. Roy. Astron. Soc.,* **405**, Is. 2, 995—1005. https://doi.org/10.1111/j.1365-2966.2010.16546.x
53. Pulatova N. G.; Vavilova I. B.; Sawangwit U., et al. (2015). The 2MIG isolated AGNs - I. General and multiwavelength properties of AGNs and host galaxies in the northern sky. *Mon. Notic. Roy. Astron. Soc.,* **447**, Is. 3, 2209—2223. https://doi.org/10.1093/mnras/stu2556
54. Pulatova N. G., Vavilova I. B., Vasylenko A. A., Ulyanov O. M. (2023). Radio properties of the low-redshift isolated galaxies with active nuclei. *Kinemat. Phys. Celest. Bodies*, **39**, Is. 2, 98—115. https://doi.org/10.3103/S088459132302006X
55. Sabater J., Best P. N., Heckman T. M. (2015). Triggering optical AGN: the need for cold gas, and the indirect roles of galaxy environment and interactions. *Mon. Notic. Roy. Astron. Soc.,* **447**, Is. 1, 110—116. https://doi.org/10.1093/mnras/stu2429
56. Sabater J., Leon S., Verdes-Montenegro L., et al. (2008). The AMIGA sample of isolated galaxies. VII. Far-infrared and radio continuum study of nuclear activity. *Astron. and Astrophys.,* **486**, Is. 1, 73—83. https://doi.org/10.1051/0004-6361:20078785
57. Schawinski K., Thomas D., Sarzi M., et al. (2007). Observational evidence for AGN feedback in early-type galaxies. *Mon. Notic. Roy. Astron. Soc.,* **382**, Is. 4, 1415—1431. https://doi.org/10.1111/j.1365-2966.2007.12487.x
58. Soszyński I., Gieren W., Pietrzyński G., et al. (2006). The Araucaria Project: Distance to the Local Group Galaxy NGC 3109 from Near-Infrared Photometry of Cepheids. *Astrophys. J.*, **648**(1), 375—382. https://doi.org/10.1086/505789
59. Toomre A. (1964). On the gravitational stability of a disk of stars. *Astrophys. J.*, **139**, 1217—1238. https://doi.org/10.1086/147861
60. Vavilova I. B.; Dobrycheva D. V., Khramtsov V., et al. (2024). Machine Learning of Galaxy Classification by their Images and Photometry. *Publ. ASP*, **535**, 103—107.
61. Vavilova I. B., Dobrycheva D. V., Vasylenko M. Yu. (2021). Machine learning technique for morphological classification of galaxies from the SDSS. I. Photometry-based approach. *Astron. and Astrophys.*, **648**, id. A122, 1—14. https://doi.org/10.1051/0004-6361/202038981
62. Vavilova I. B., Elyiv A. A., Dobrycheva D. V., et al. (2021). Voronoi tessellation method in astronomy. *Intelligent Astrophysics*. Springer, Cham, 57—79. ISBN: 978-3-030-65867-0. https://doi.org/10.1007/978-3-030-65867-0_3
63. Vavilova I. B., Fedorov P. M., Dobrycheva D. V., et al. (2024). An advanced approach to the definition of the “Milky Way galaxies-analogues”. *Space Sci. and Technol.*, **30**, Is. 4, 81—90. https://doi.org/10.15407/knit2024.04.081
64. Vavilova I. B., Ivashchenko G. Yu., Babyk Iu. V., et al. (2015) The astrocosmic databases for multi-wavelength and cosmological properties of extragalactic sources. *Space Sci. and Technol.*, **21**, Is. 5, 94—107. https://doi.org/10.15407/knit2015.05.094
65. Vavilova I. B., Khramtsov V., Dobrycheva D. V., et al. (2022). Machine learning technique for morphological classification of galaxies from SDSS. II. The image-based morphological catalogs of galaxies at $0.02 < z < 0.1$. *Space Sci. and Technol.*, **28**, Is. 1, 3—22. https://doi.org/10.15407/knit2022.01.003
66. Vavilova I., Kompaniiets O., Vasylenko A., et al. (2025). Milky Way analogues as the isolated AGNs: multiwavelength data incompleteness. *ASP Conf. Ser.*, ISSN (ISSN-L) 1050-3390.
67. Wang P., Kang X., Libeskind N. I., et al. (2020). A robust determination of halo environment in the cosmic field. *New Astronomy*, **80**, id. 101405. https://doi.org/10.1016/j.newast.2020.101405

*О. В. Компанієць*[1], мол. наук. співроб.
https://orcid.org/0000-0002-8184-6520
E-mail: kompaniets@mao.kiev.ua
*І. Б. Вавилова*[1], зав. відділу, д-р фіз.-мат. наук, проф.
https://orcid.org/0000-0002-5343-1408
*О. М. Кухар*[1], мол. наук. співроб., аспірант
https://orcid.org/0009-0000-7042-9214
*Д. В. Добричева*[1], старш. наук. співроб., канд. фіз.-мат. наук
https://orcid.org/0000-0001-5557-3453
*П. М. Федоров*[2], зав. лаб., д-р фіз.-мат. наук
https://orcid.org/0000-0001-6745-3731
*А. М. Дмитренко*[1], наук. співроб., канд. фіз.-мат. наук
https://orcid.org/0000-0001-9380-6641
*В. П. Храмцов*[2], інженер-програміст
https://orcid.org/0000-0003-1744-7071
*О. М. Сергієнко*[1], старш. наук. співроб., канд. фіз.-мат. наук
https://orcid.org/0000-0002-9212-7118
*А. А. Василенко*[1], старш. наук. співроб., канд. фіз.-мат. наук
https://orcid.org/0000-0001-7142-6576

## ГАЛАКТИКИ-АНАЛОГИ ЧУМАЦЬКОГО ШЛЯХУ ТА ГАЛАКТИКИ З БАРОМ: ЩІЛЬНІСТЬ ОТОЧЕННЯ У МІСЦЕВОМУ ВСЕСВІТІ

Щільність середовища, у якому розташовані галактики в межах космічної павутини, дає змогу визначити їхні тривимірні положення у філаментах, войдах, групах і скупченнях великомасштабної структури Всесвіту. Цей параметр відображає розподіл баріонної речовини та вплив гало темної матерії на еволюцію галактик. Розуміння параметрів щільності середовища є ключовим для виявлення зовнішніх фізичних процесів — таких як зворотний зв'язок від наднових і активних ядер галактик, припливні взаємодії, тиск від міжгалактичного середовища та великомасштабні потоки речовини, — які визначають властивості галактик поза їхніми внутрішніми характеристиками.

У цій статті ми зосередилися на ізольованих галактиках із барами як на сукупності галактик, до якої належать аналоги галактики Чумацький Шлях. Для отримання параметрів локальної щільності середовища та перевірки критерію ізольованості ($|\Delta v| \leq 500$ км/с) розроблено Python-пакет, що працює у двох режимах за червоним зсувом: для близьких ($z_0 < 0.02$) та віддалених ($z_0 \geq 0.02$) галактик. Локальні щільності $\Sigma$ визначено методами найближчих $k$-сусідів і тесселяції Вороного та класифіковано за чотирма категоріями: войд ($\Sigma < 0.05$), філамент ($0.05 \leq \Sigma < 0.5$), група ($0.5 \leq \Sigma < 2.0$) і скупчення ($\Sigma \geq 2.0$).

Вибірка складається з 311 ізольованих барованих галактик із каталогу 2MIG, доповнених системами-аналогами Чумацького Шляху ($z < 0.07$), що охоплюють північну півкулю неба. Виявлено 157 галактик у войдах, 84 у філаментах, 27 у групах та 11 у скупченнях; 30 об'єктів не мають виявлених сусідів. Підвибірка з 67 галактик перебуває в областях надзвичайно низької щільності ($\Sigma_3$D < 0.01 gal Mpc$^{-3}$), тоді як 22 галактики мають найближчого сусіда на відстані понад 5 Мпк, що свідчить про їх розташування у великих космологічних войдах. Галактика Чумацький Шлях ($\Sigma_{5NN} \approx$ $\approx 0.13$ gal Mpc$^{-3}$, $R \approx 2.1$ Мпк) та її близький аналог NGC 3521 розташовані у філаментному середовищі, характерному для проміжних щільностей на межі сусіднього войду. Для галактик із $z > 0.02$ локальні оцінки щільності дали змогу виділити додаткових кандидатів-аналогів Чумацького Шляху: CGCG 208-043 (за оцінками 3D-тесселяції Вороного) та NGC 5231 і CGCG 047-026 (за методом п'ятого найближчого сусіда). Отримані результати підкреслюють важливість локальної щільності середовища як одного з ключових параметрів для ідентифікації та характеристики галактик-аналогів Чумацького Шляху.